# Distinguishing differential susceptibility, diathesis-stress and vantage sensitivity: beyond the single gene and environment model

**Alexia Jolicoeur-Martineau[1], Jay Belsky[2], Eszter Szekely[1,3], Keith F. Widaman[4], Michael Pluess[5], Celia Greenwood[1,6] and Ashley Wazana[1,3,7*].**

[1]Lady Davis Institute for Medical Research, Jewish General Hospital, Montreal, Qc, Canada; [2]Department of Human Ecology, University of California, Davis, USA; [3]Department of Psychiatry, Faculty of Medicine, McGill University, Montreal, Qc, Canada; [4]Graduate School of Education, University of California, Riverside, USA; [5]Department of Biological and Experimental Psychology, Queen Mary University of London, UK; [6]Ludmer Centre for Neuroinformatics and Mental Health, Montreal, Qc, Canada; [7]Douglas Mental Health University Institute, Montreal, Qc, Canada.

* Corresponding author

E-mail: ashley.wazana@mcgill.ca
Address correspondence to Ashley Wazana, Centre for Child Development and Mental Health, Jewish General Hospital, 4335 Cote Sainte Catherine Road, Montreal, Quebec, H3T 1E4 Montreal, Quebec, Canada; email: ashley.wazana@mcgill.ca

## Abstract

Currently, two main approaches exist to distinguish differential susceptibility from diathesis-stress and vantage sensitivity in genotype × environment interaction (G×E) research: Regions of significance (RoS) and competitive-confirmatory approaches. Each is limited by their single-gene/single-environment foci given that most phenotypes are the product of multiple interacting genetic and environmental factors. We thus addressed these two concerns in a recently developed R package (LEGIT) for constructing G×E interaction models with latent genetic and environmental scores using alternating optimization. Herein we test, by means of computer simulation, diverse G×E models in the context of both single and multiple genes and environments. Results indicate that the RoS and competitive-confirmatory approaches were highly accurate when the sample size was large, whereas the latter performed better in small samples and for small effect sizes. The competitive-confirmatory approach generally had good accuracy (a) when effect size was moderate and $N \geq 500$ and (b) when effect size was large and $N \geq 250$, whereas RoS performed poorly. Computational tools to determine the type of G×E of multiple genes and environments are provided as extensions in our LEGIT R package.

# Introduction

## Genotype × environment interactions

Over the past 15 years, several distinct conceptual models of how individual and environmental characteristics interact in shaping development have been used in studying and interpreting evidence of genotype × environment interaction (G×E). These models include (a) *diathesis-stress*, stipulating that some individuals carry "risk" genes that make them disproportionately susceptible to adverse environmental conditions (Zubin & Spring, 1977); (b) *vantage sensitivity*, stipulating that some individuals disproportionately benefit from supportive conditions due to their genetic make-up (Pluess & Belsky, 2013); and (c) *differential susceptibility*, stipulating that some individuals are more developmentally plastic "for better and for worse" (Belsky, Bakermans-Kranenburg & van IJzendoorn, 2007), being more susceptible to both negative effects of adversity and beneficial effects of support (Belsky, 1997a, 1997b), thus reconceptualising would-be "risk" genes into more general "plasticity" or "sensitivity" genes. Weak and strong versions of each model can be envisioned; in strong models, some individuals are affected by the environmental exposure of interest while others are not, whereas in weak models all are affected by the environmental exposure, but some more strongly than others. See Figure 1 for a graphical representation of the a) strong and b) weak models.

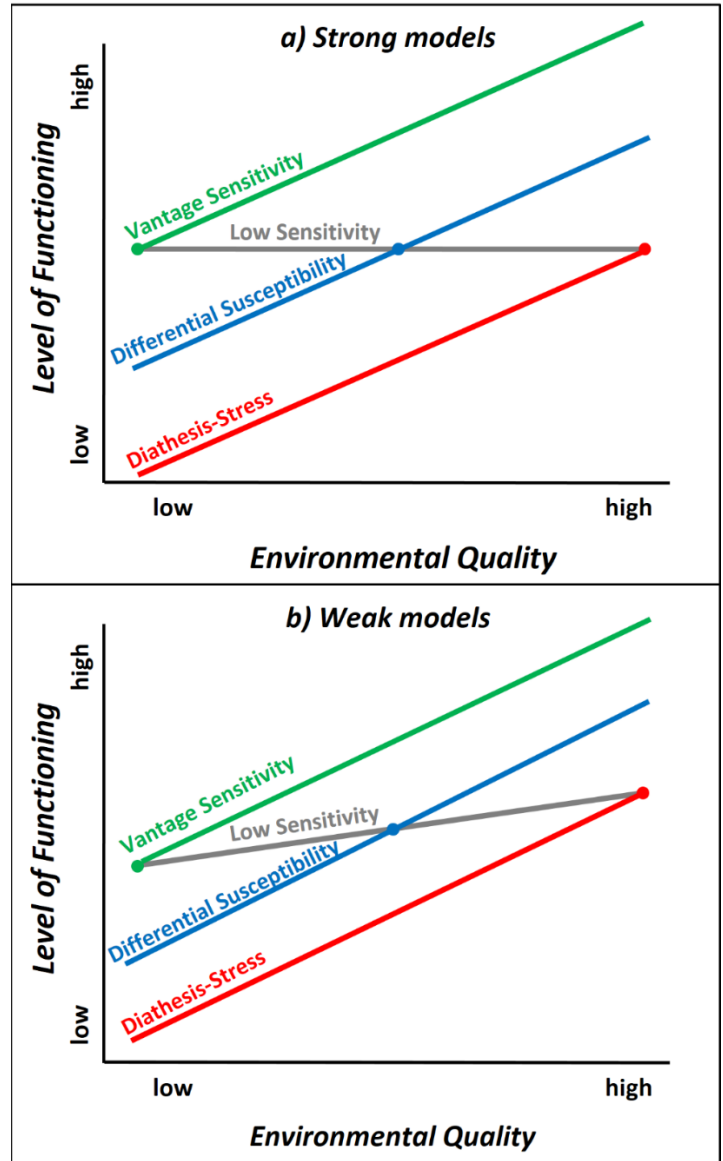

**Figure 1**. *A graphical representation of the three G×E hypotheses tested (vantage sensitivity, differential susceptibility, and diathesis-stress) and where the crossover point is situated in each case assuming a) weak and b) strong models.*

Most work evaluating diverse models of G×E interaction has been exploratory in character, failing to formally evaluate which model fit the data best (see Assary, Vincent, Keers, and Pluess [2017] for a recent review of the subject). Kochanska, Kim, Barry, and Philibert (2011) were the first to use the now classical regions of significance (RoS) analysis (Aiken, West, & Reno, 1991; Hayes & Matthes, 2009; Preacher, Curran, & Bauer, 2006) to differentiate diathesis-stress from differential susceptibility. Assuming a single binary genetic variant (e.g., short vs. long 5-HTTLPR allele), the RoS approach determines the range of values of the environment where the environment-predicting-outcome regression lines significantly differ from each other. A G×E model is considered to reflect diathesis-stress when the lines are significantly different only in the lower observable range of the environmental quality measure (e.g., poor parenting), thereby reflecting adversity; vantage sensitivity when the lines are significantly different only in the upper observable range of the environmental quality measure (e.g., good parenting), thereby reflecting support of enrichment; differential-susceptibility when the lines are significantly separated at both ends of the environmental quality measure. See Figure 2 for a graphical representation of the a) diathesis-stress model, b) differential susceptibility model, and c) vantage sensitivity model with the RoS.

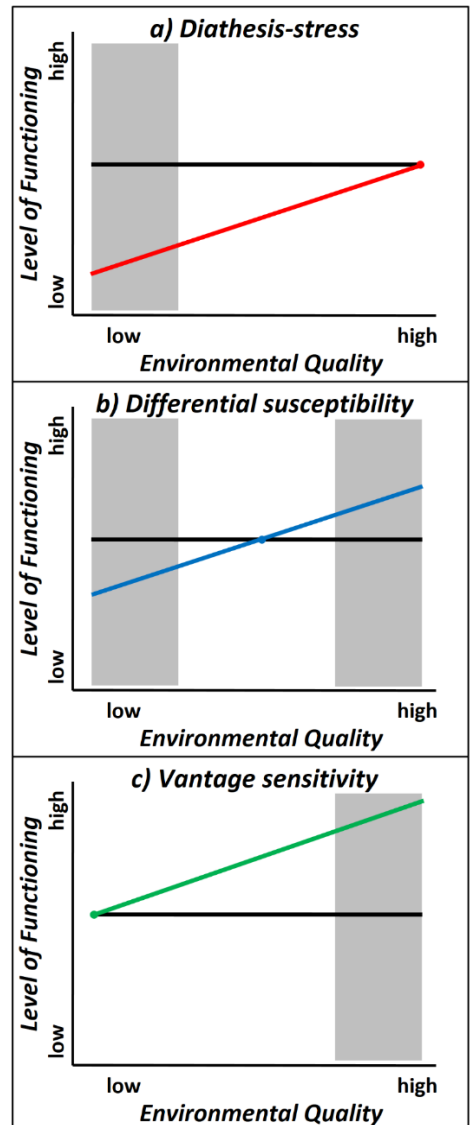

**Figure 2.** *A graphical representation of the three G×E hypotheses tested (vantage sensitivity, differential susceptibility, and diathesis-stress) and the regions of significance (in grey; area where the slopes are significantly different from one another).*

In seeking to advance new methods for empirically distinguishing different conceptual models of G×E interaction, Roisman et al. (2012) highlighted several limitations of the RoS approach. Because statistical significance is a function of sample size, larger studies would be more likely to detect significantly different slopes in both the lower and upper end of the spectrum than smaller ones. Additionally, the RoS approach is likely sensitive to Type 1 errors, particularly when testing multiple variants and environments. For example, if one has ten genetic variants and five environmental measures, one could conduct 10*5 = 50 pairwise tests. Assuming testing at the 5% level (i.e. p-value < .05; 95% confidence interval), the probability of having at least one false positive out of 50 tests is 92.3%. Such multiple testing means that the confidence intervals should be adjusted to allow for the consideration of multiple hypotheses in the same empirical effort. However, in practice, many researchers do not adjust for multiple testing, let alone report their non-significant results (Earp & Trafimow, 2015; Simmons, Nelson, & Simonsohn, 2011). Finally, as a result of non-linearity, the slopes might never cross even when they are significantly different given the observable range of the environmental variable.

About the same time that Roisman et al. (2012) were expressing the concerns just raised, Widaman and associates (2012; Belsky et al., 2013) proposed a method for competitively evaluating the fit of diathesis-stress and differential-susceptibility models of G×E, one which was confirmatory rather than exploratory. In their competitive-confirmatory approach, the cross-over point is the critical parameter. Whereas the diathesis-stress model is constructed by fixing the crossover point at the higher end of the environmental quality spectrum, the differential-susceptibility model is constructed by treating the cross-over point as an unknown parameter to be estimated. A simple adjustment to the Widaman et al. approach would also enable it to evaluate vantage sensitivity—by fixing the crossover point at the lower end of the environmental quality spectrum. Criteria such as AIC or BIC can be used to determine which alternative model fits the data best. Notably, this competitive-confirmatory approach still relies indirectly on hypothesis-testing, but only for the purpose of rejecting models of differential susceptibility which have a confidence interval for the cross-over point that lies outside the observable range. To date, these models have been used principally in studies focused on a single candidate gene and a single environmental exposure.

**Multiple genes and environments**

As mentioned by Roisman et al. (2012), the risk of Type 1 errors in G×E testing approaches is high considering that one needs to run separate analyses for each individual pair of gene-environment variables. This problem is even more relevant and pronounced in the era of big data when the cost of DNA sequencing has rapidly decreased and became more affordable in large-scale epidemiological research (Wetterstrand, 2016). Another problem with single-gene and single-environment G×E models is that most G×E models involving single genes and single environments have very small effect sizes and low replication rates (P. H. Lee et al., 2012; Risch et al., 2009). This is not surprising based on the common understanding that most phenotypes are affected by multiple genes, to say nothing of multiple environments.

These considerations highlight the need to develop G×E approaches that can handle multiple genes and/or multiple environments—which was in fact the reason we developed an approach using alternating optimization (Jolicoeur-Martineau et al., 2017) to construct a genetic score (*g*) × environmental score (*e*) model, in which *g* is a weighted sum of the genetic variants and *e* is a weighted sum of the environments. This approach estimates, in turn, the weights of the G×E model, the weights of the genetic score and the weights of the environmental score. We call the models fitted using this approach "Latent Environmental and Genetic InTeraction" (LEGIT) models. With this approach, instead of testing the interaction of each gene/environment pair (n_genes × n_env tests), one has to test only *once* for the presence of a genetic score × environmental score interaction. How much each genetic/environmental variable contribute to their corresponding genetic/environmental score can then be tested for each individual variables (n_genes + n_env tests), however, this latter is not required to test for the presence of a significant G×E. We recommend that instead of making a decision based on p-values of the G×E term, one should rather compare the BIC of the models with and without an interaction term; doing so renders false positives very unlikely (as is shown in the results section).

We wish to emphasize the distinction between (a) testing of multiple genes and environments in serial models, and (b) the use of a model which fits, within set parameters of genes and environments established a priori, the optimal weights for each term of the equation. The vast majority of statistical models used in the developmental-behavioral sciences (e.g. mixed models, logistic regression, and structural equation models) use the latter, with an iteration-based optimization approach to estimate the parameters. Thus, our alternating optimization is no different from analytic approaches commonly used in our field of research.

Here we seek to extend this initial work by adapting and applying RoS and the competitive-confirmatory approaches within LEGIT models to test for the pattern of interaction between the genetic and environmental score. This affords the differentiation of differential susceptibility, diathesis-stress, and vantage sensitivity in a multi-gene/multi-environment G×E setting.

In an effort to extend work evaluating alternative G×E models, we evaluate, via simulation, the relative accuracy of the RoS and competitive-confirmatory approaches under a wide range of scenarios; these include single and multiple genes, single and multiple environmental measures, different effect sizes, different sample sizes, symmetric and skewed distributions of the environment and crossover points at 25% or 50% of the measured environmental quality range.

Below, we first briefly review the RoS, competitive-confirmatory, and alternating optimization approaches. Next, we discuss how to combine the RoS and competitive-confirmatory approaches with alternating optimization. Finally, we use statistical simulations to test whether:

1) Confirmatory and competitive testing has differential accuracy in determining the pattern of G×E for small sample sizes and effect sizes compared to RoS.
2) Our method combining competitive-confirmatory and alternating optimization approaches has good accuracy in a multi G×E model (i.e., 4 genetic variants and 3 environmental measures) assuming (1) a moderate effect size and large sample size (i.e., N=500 or greater) or (2) a large effect size and small to moderate sample size (i.e., N=250 or greater).

## Methods

### G×E model definition

A standard G×E model can be represented in the following way:

$$\boldsymbol{y} = \beta_0 + \beta_e \boldsymbol{e} + \beta_g \boldsymbol{g} + \beta_{eg} \boldsymbol{eg} + \boldsymbol{\varepsilon},$$ 

(1)

where $\boldsymbol{y}$ is a vector representing the $n$ observed outcomes, $\beta_0$ is the intercept, $\beta_e$ the regression weight for the environment main effect ($\boldsymbol{e}$), $\beta_g$ the regression weight for the gene main effect ($\boldsymbol{g}$), and $\beta_{eg}$ is the regression weight for the product of environment and genes ($\boldsymbol{eg}$) which represents the G×E interaction. For a fitted model, we use the following formulation:

$$E\left[\boldsymbol{y} \mid \boldsymbol{e}, \boldsymbol{g}\right] = \hat{\beta}_0 + \hat{\beta}_e \boldsymbol{e} + \hat{\beta}_g \boldsymbol{g} + \hat{\beta}_{eg} \boldsymbol{eg}. \tag{2}$$

The hat ($\wedge$) is used on top of parameters to represent that these are estimated parameters rather than the original ones. This notation will be used through the paper.

**Simple slopes and regions of significance**

Assuming a single binary genetic variant (e.g., short vs. long 5-HTTLPR allele), a G×E model has two environment-predicting-outcome regression lines: one representing individuals with, putatively, low environmental sensitivity and another representing individuals with, putatively, high environmental sensitivity. The traditional approach for distinguishing between differential susceptibility, diathesis-stress, and vantage sensitivity is to determine at which values of the environment the lines differ significantly from each other.

To this end, one must first reparametrize a G×E model as an explicit function of the genotype:

$$E\left[\boldsymbol{y} \mid \boldsymbol{e}, \boldsymbol{g}\right] = \hat{\beta}_0 + \hat{\beta}_e \boldsymbol{e} + \hat{\beta}_g \boldsymbol{g} + \hat{\beta}_{eg} \boldsymbol{eg}$$

$$= \left(\hat{\beta}_0 + \hat{\beta}_e \boldsymbol{e}\right) + (\hat{\beta}_g + \hat{\beta}_{eg} \boldsymbol{e}) \boldsymbol{g}$$

$$= \hat{w}_0 + \hat{w}_1 \boldsymbol{g}.$$

We call $\hat{w}_0$ the simple intercept and $\hat{w}_1$ the simple slope. The simple slope represents the slope as a function of $\boldsymbol{g}$. This means that if $\hat{w}_1$ is significantly different from zero, then the slopes for different values of $\boldsymbol{g}$ (e.g. the slopes for $\boldsymbol{g} = 0$ and $\boldsymbol{g} = 1$ if $\boldsymbol{g}$ is dichotomous) are significantly different from one another.

The variance of the simple slope is:

$$Var\left[\hat{w}_1 \mid \boldsymbol{e}\right] = Var\left(\hat{\beta}_g\right) + Var\left(\hat{\beta}_{eg} \boldsymbol{e}\right) + 2Cov\left(\hat{\beta}_g, \hat{\beta}_{eg} \boldsymbol{e}\right)$$

$$= Var\left(\hat{\beta}_g\right) + \boldsymbol{e}^2 Var\left(\hat{\beta}_{eg}\right) + 2\boldsymbol{e} Cov\left(\hat{\beta}_g, \hat{\beta}_{eg}\right).$$

Using standard regression asymptotic theory (Aiken et al., 1991), it can be shown that:

$$\frac{\hat{w}_1}{\sqrt{Var\left[\hat{w}_1 \mid \boldsymbol{e}\right]}} \sim t_{N-5}$$

Where $t_{N-5}$ is a Student's t-distribution with $N$-5 degrees of freedom (df) and $N$ is the number of observations.

Based on the above equations, it is possible to test whether the simple slope is significantly different from zero, and thus determine if the slopes for different values of $\boldsymbol{g}$ are significantly different from one another at fixed values of $\boldsymbol{e}$. However, this does not tell us at which values of $\boldsymbol{e}$ is $\hat{w}_1$ significant. The Johnson-Neyman (J-N) technique (Johnson & Fay, 1950) uses this formula, in a backward-fashion, to find the range of values of $\boldsymbol{e}$ where the simple slope is significantly different from zero. This range corresponds to the regions of significance (RoS) and it can be defined by a lower bound L and an upper bound U. When $e$ is larger than L and smaller than U, the slopes for different values of $g$ do not differ significantly, whereas outside these bounds they do.

Kochanska et al. (2011) suggested using RoS (Aiken et al., 1991; Hayes & Matthes, 2009; Preacher et al., 2006) to help differentiate diathesis-stress from differential susceptibility. When the regions of significance are within the observable ranges of values for $e$, the data are considered to reflect differential susceptibility. When only the lower bound L of the RoS is within the observable range, the data is considered to reflect diathesis-stress. When only the upper bound U of the RoS is within the observable range, the data is considered to reflect vantage sensitivity. If neither L nor U is within the observable range, results are inconclusive.

## Confirmatory and competitive models

As shown above, the RoS approach focused entirely on the lower and upper bound of the regions of significance. An alternative point of view is to focus instead on the crossover point, the point where the slopes for the different values of the environment cross. It was shown (Aiken et al., 1991; Widaman et al., 2012) that the crossover can be found using the following formula:

$$c = -\frac{\beta_g}{\beta_{eg}}. \tag{4}$$

Similarly to the RoS approach, if the complete 95% confidence interval of the crossover point is within the observable range of the environment, it is suggestive of differential susceptibility. To obtain a confidence interval for the crossover point, Widaman et al. (2012) suggested fitting a model where the crossover point $c$ is an unknown parameter to be estimated. The authors showed that the standard G×E formulation (1) can be reparametrized as:

$$y = \beta_0 + \beta_e(e - c) + \beta_{eg}(e - c)g + \varepsilon. \tag{5}$$

Although $\beta_g$ does not appear in this equation, Widaman et al. (2012) have shown that this parametrization is equivalent due to the inclusion of the crossover point $c$.

Using nonlinear regression estimation, model (5) can be fitted and the crossover point can be estimated directly along with its confidence intervals. It is important to note that the estimate of the crossover point is not normally distributed (S. Lee, Lei, & Brody, 2015; Marsaglia, 1965), suggesting that the resulting confidence interval may be biased. However, it has been shown that when the sample size is large enough (N > 500), the crossover point is approximately normally distributed (S. Lee et al., 2015; Marsaglia, 1965). In addition, even when the sample size is small, the non-normality of the crossover point does not lead to significant problems in the case of the competitive-confirmatory approach, as it does not rely exclusively on confidence intervals to distinguish between differential susceptibility and diathesis-stress; this is in contrast to the RoS approach. Instead, Widaman et al. (2012) constructed confirmatory and competitive models of differential susceptibility and diathesis-stress to determine which model best fits the data. In accordance, the purpose of the confidence interval of the crossover point is only to verify that the crossover point is within the observable range of the environment when a differential susceptibility model is deemed to be the best fit-wise. To prevent any confusion, we will use the notation $c$ for the true crossover point, $\hat{c}$ for the estimated crossover of the differential susceptibility models, $c_{low}$ for the crossover point of the vantage sensitivity models and $c_{high}$ for the crossover point of the diathesis-stress models.

Although the original paper by Widaman et al. (2012) focused only on the diathesis-stress and differential susceptibility models, we extend this approach in the current effort by also presenting the vantage sensitivity models. Furthermore, we extend all concepts to situations where multiple environmental variables contribute to an environmental score, and where multiple genetic variables contribute to a genetic score. Accordingly, there are six models tested in our extension to multidimensional situations:

1) weak vantage sensitivity ($\beta_e$ estimated, $c_{low} = \min(\boldsymbol{e})$)
2) strong vantage sensitivity ($\beta_e = 0$, $c_{low} = \min(\boldsymbol{e})$)
3) weak differential susceptibility ($\beta_e$ estimated, $c$ estimated)
4) strong differential susceptibility ($\beta_e = 0$, $c$ estimated)
5) weak diathesis-stress ($\beta_e$ estimated, $c_{high} = \max(\boldsymbol{e})$)
6) strong diathesis-stress ($\beta_e = 0$, $c_{high} = \max(\boldsymbol{e})$)

All six models are constructed using equation (5). The models of differential susceptibility [3] and 4)] need to estimate an extra parameter, namely, the crossover point; therefore, Widaman et al. (2012) suggested using the AIC (Akaike, 1998) or the BIC (Schwarz, 1978) to evaluate the quality of fit.

**Latent genetic and environmental score models**

Instead of considering a linear regression on a single observed genetic variant and environment, as in equation (1), Jolicoeur-Martineau et al. (2017) instead considered a linear regression on latent genetic ($\boldsymbol{g}$) and environmental ($\boldsymbol{e}$) scores and their interaction. These latent scores are defined as the weighted sum of their corresponding variables (genetic or environmental). The goal thus becomes not only to estimate the parameters of the G×E, but also to estimate the weights of the genetic and environmental variables that make up the latent variables. Jolicoeur-Martineau et al. (2017) defined their model in three parts: the genetic score $\boldsymbol{g}$, the environmental score $\boldsymbol{e}$, and the main model.

$$\boldsymbol{g} = \sum_{j=1}^{k} p_j \boldsymbol{g}_j$$

$$\boldsymbol{e} = \sum_{l=1}^{s} q_l \boldsymbol{e}_l \qquad (6)$$

$$y = \beta_0 + \beta_e \boldsymbol{e} + \beta_g \boldsymbol{g} + \beta_{eg} \boldsymbol{eg} + \varepsilon,$$

where $\boldsymbol{p} = (p_1, p_2, \ldots, p_k)$ is a vector of parameters for the k genetic variables $\boldsymbol{g}_1, \boldsymbol{g}_2, \ldots, \boldsymbol{g}_k$ and $\boldsymbol{q} = (q_1, q_2, \ldots, q_s)$ is a vector of parameters for the s environmental variables $\boldsymbol{e}_1, \boldsymbol{e}_2, \ldots, \boldsymbol{e}_s$.

Note that there are infinitely many possibilities for $\boldsymbol{p}$ and $\boldsymbol{q}$ that lead to the same model. For example, $\beta_e$, $\beta_{eg}$ with $\boldsymbol{Cp}$, where $C$ is a constant, leads to the exact same model as $C\beta_e$, $C\beta_{eg}$ with $\boldsymbol{p}$:

$$y = \beta_0 + \beta_e \boldsymbol{e} + \beta_g \sum_{j=1}^{k}(Cp_j)\boldsymbol{g}_j + \beta_{eg}\boldsymbol{e}\sum_{j=1}^{k}(Cp_j)\boldsymbol{g}_j + \boldsymbol{\varepsilon}$$

$$= \beta_0 + \beta_e \boldsymbol{e} + C\beta_g \sum_{j=1}^{k}p_j\boldsymbol{g}_j + C\beta_{eg}\boldsymbol{e}\sum_{j=1}^{k}p_j\boldsymbol{g}_j + \boldsymbol{\varepsilon}$$

$$= \beta_0 + \beta_e \boldsymbol{e} + (C\beta_g)\boldsymbol{g} + (C\beta_{eg})\boldsymbol{eg} + \boldsymbol{\varepsilon}.$$

To prevent infinite possibilities for $\boldsymbol{p}$ and $\boldsymbol{q}$, the following restrictions are added:

$$\sum_{j=1}^{k}\left|p_j\right| = 1,$$

$$\sum_{l=1}^{s}\left|q_l\right| = 1.$$

(7)

This restriction also aids interpretation because the weights of the genetic and environmental scores represent the relative contribution of the individual genetic variants and environmental variables to their respective composites.

   To estimate the weights of the genetic score, the environmental score, and the main models, an alternating optimization approach is used. We first initialize $\boldsymbol{p}$ to $(1/k, 1/k, \ldots, 1/k)$ and $\boldsymbol{q}$ to $(1/s, 1/s, \ldots, 1/s)$, i.e. we assume that all genetic and environmental variables are equally weighted; this is equivalent to taking the average of all the genetic or environmental variables. Then, the approach comprises three steps (1) estimating $\beta_0, \beta_e, \beta_g, \beta_{eg}$ assuming that $\boldsymbol{p}$ and $\boldsymbol{q}$ are known, (2) estimating $\boldsymbol{p}$ assuming that $\boldsymbol{q}$ and $\beta_0, \beta_e, \beta_g, \beta_{eg}$ are known, and (3) estimating $\boldsymbol{q}$ assuming that $\boldsymbol{p}$ and $\beta_0, \beta_e, \beta_g, \beta_{eg}$ are known (Jolicoeur-Martineau et al., 2017). This is done in iterative steps until convergence, at each step ensuring that the parameters of the genetic and environmental scores sum to one in absolute values (constraints from equation (7)). Table 1 shows an example of alternating optimization being used to estimate a LEGIT G×E model with four genes and one environmental variable. The approach is discussed in greater details in our previous methodological paper (Jolicoeur-Martineau et al., 2017).

**Table 1:** *Example of alternating optimization being used to estimate a simple G×E model with 4 genetic variables and 1 environment variable; $E[\mathbf{y}] = \beta_0 + \beta_g \boldsymbol{g} + \beta_e \boldsymbol{e} + \beta_{eg} \boldsymbol{eg}$ with a single e and $\boldsymbol{g} = p_1 g_1 + p_2 g_2 + p_3 g_3 + p_4 g_4$.*

| Parameters<br><br>Step | $\beta_0$ | $\beta_g$ | $\beta_e$ | $\beta_{eg}$ | $p_1$ | $p_2$ | $p_3$ | $p_4$ | $R^2$ |
|---|---|---|---|---|---|---|---|---|---|
| 0 - initialization | ? | ? | ? | ? | **.25** | **.25** | **.25** | **.25** | ? |
| 1 - main | **5.12** | **-.77** | **3.05** | **-.90** | .25 | .25 | .25 | .25 | .843 |
| 1 - genes | 5.12 | -.77 | 3.05 | -.90 | **.87** | **-.47** | **1.56** | **-.93** | .947 |
| $\sum_{j=1}^{k}\left\|p_j\right\| = 1$ | 5.12 | **-2.96** | 3.05 | **-3.46** | **.23** | **-.12** | **.41** | **-.24** | .947 |
| 2 - main | **4.99** | **-1.83** | **3.05** | **-3.84** | .23 | -.12 | .41 | -.24 | .952 |
| 2 - genes | 4.99 | -1.83 | 3.05 | -3.84 | **.23** | **-.13** | **.40** | **-.25** | .952 |
| $\sum_{j=1}^{k}\left|p_j\right|=1$ | 4.99 | **-1.86** | 3.05 | **-3.90** | **.22** | **-.13** | **.40** | **-.25** | .952 |
| ... | ... | ... | ... | ... | ... | ... | ... | ... | ... |
| End | 4.95 | -1.87 | 2.96 | -3.91 | .21 | -.14 | .39 | -.26 | .953 |
| Optimal | 5 | -2 | 3 | -4 | .20 | -.15 | .40 | -.25 | |

Implementations in R and SAS are freely available online on CRAN (cran.r-project.org/web/packages/LEGIT) and Github (github.com/AlexiaJM/LEGIT).

**Combining RoS with alternating optimization**

Combining RoS with alternating optimization is a simple step, whereby one can apply RoS directly to the main model of equation (6) assuming that $p$ and $q$ are known.

**Combining confirmatory with alternating optimization**

To combine the competitive-confirmatory approach with alternating optimization, model (6) can be reparametrized and reformulated to include a crossover point, thus enabling the testing of the interaction, in the following way:

$$g = \sum_{j=1}^{k} p_j g_j$$

$$e = \sum_{l=1}^{s} q_l e_l$$

$$y = \beta_0 + \beta_e(e-c) + \beta_g g + \beta_{eg}(e-c)g + \varepsilon,$$

(8)

This new formulation corresponds to the competitive-confirmatory model (Belsky, Pluess, & Widaman, 2013; Widaman et al., 2012) shown in equation (5). Using this formulation, estimating the crossover point $c$ would require the use of a nonlinear regression; however, by simply adding a negative intercept to the environmental score, this equation can be reparametrized in a way that each part remains a linear model, as shown below:

$$g = \sum_{j=1}^{k} p_j g_j$$

$$e' = -c + \sum_{l=1}^{s} q_l e_l$$

$$y = \beta_0 + \beta_e e' + \beta_g g + \beta_{eg} e' g + \varepsilon,$$

(9)

Accordingly, the alternating optimization approach can be easily modified to include a crossover point and test for the 6 different hypotheses mentioned above (i.e., diathesis-stress, differential susceptibility or vantage sensitivity; weak or strong). See Appendix A for more details on how to adapt the alternating optimization algorithm for competitive-confirmatory testing and Appendix B for special considerations regarding testing for the presence of a G×E in case of multiple genes and environments.

## Simulation set-up

We have proposed extensions of the concepts of RoS and competitive-confirmatory approaches to multidimensional environmental and genetic scores. To test the performance of these extensions with one or multiple genes and environments, we examined how often these two approaches can correctly determine the pattern of interaction (i.e., diathesis-stress, vantage sensitivity, or differential susceptibility) under various scenarios. We simulated 100 samples for each of the six different models (representing the three hypotheses, each as weak or strong) and report the average accuracy of all models. We defined accuracy as the percentage of correctly assigned patterns of interactions. Additionally, we simulated 100 samples of the model without a G×E (genes and environments only) to estimate the false positive rate (i.e. the percentage of models not assigned as having no evidence of G×E when there is no actual G×E).

We present one scenario with a single genetic variant and single environment (traditional G×E model) and one scenario with four genetic variants and three environmental variables. In both scenarios, these variables are sampled from the following distributions:

$$g_j \sim \text{Bernouilli}(.30),$$
$$e_l \sim \text{Beta}(2, \beta).$$

Given that the vast majority of environmental measures in psychological/epidemiological research are ordinal (e.g. Likert scale) and thus bounded by a minimum and maximum value, we chose a beta distribution for the environmental factors. Note that the beta distribution is bounded by 0 and 1. We present two scenarios, one with $\beta = 2$ which leads to a fully symmetric normal-like distribution and one with $\beta = 4$ for a highly left-skewed distribution. Left-skewed variables are very frequently used as environmental factors (e.g., socioeconomic status, income, depressive symptoms, etc.) which justifies using this type of distribution. A priori, we theorized that it might be harder to confirm a diathesis-stress model with crossover at the maximum score of the environmental variable when very few individuals have values close to the maximum.

In all models, we let $\beta_0 = 3$ and $\beta_{eg} = 2$. The weak models have $\beta_e = 1$, while the strong models have $\beta_e = 0$. The crossover point is set to 0 for the vantage sensitivity models and 1 for the diathesis-stress models. We present two scenarios for the choice of the crossover of the differential susceptibility models: $c = .50$, the easiest case given that it is right in the middle of the observable range of the environmental score and $c = .25$, a more difficult case given that it is closer to the minimum possible value of the environmental quality score.

Assuming a Gaussian error term with variance $\sigma$ (i.e. $\varepsilon \sim \text{Normal}(0, \sigma)$), as in the case of traditional linear regression assumptions, we set up realistic simulations by using three different choices of $\sigma$ to represent small, medium, and large effect sizes. For the models with only one genetic and environmental variable, we set the variance of the error term so that the $R^2$ was .05, .10, and .15 for small, medium, and large effect sizes respectively. For the models with four genetic and three environmental variables, we set the variance of the error term so that the $R^2$ was .10, .20, and .40 for small, medium, and large effect sizes respectively. Although the values for the scenario including multiple genes and environments may seem large, they represent realistic values that have been previously observed using the alternating optimization approach (Jolicoeur-Martineau et al., 2017), because including multiple genetic variants and environmental measures in a model tends to greatly increase its predictive power. Additionally, we tested these scenarios under different sample sizes: N = 100, 250, 500, 1000, and 2000, representing very small, small, medium, large, and very large sample sizes respectively.

To summarize, we examine the following scenarios:

1. The interaction between (a) a single genetic and environmental variable or (b) 3 genetic and 4 environmental variables
2. Assuming small, medium, or large effect sizes
3. Under sample sizes: N = 100, 250, 500, 1000, or 2000
4. Assuming symmetric Beta(2,2) or skewed Beta(2,4) distribution of the environmental measures
5. Fixing the crossover of differential susceptibility at either .50 or .25

See Table 2 for a list of all the parameters used.

**Table 2:** *Parameters used in the different simulations, where E[$\mathbf{y}$] = $\beta_0$ + $\beta_e$ ($\mathbf{e}$ - c) + $\beta_{eg}$ ($\mathbf{e}$ - c)$\mathbf{g}$, for N = 100, 250, 500, 1000, 2000 and small, medium and large effect sizes.*

| Parameters<br><br>Models | $\beta_0$ | $\beta_e$ | $\beta_{eg}$ | $c$ | $\mathbf{g}$ | $\mathbf{e}$ |
|---|---|---|---|---|---|---|
| Vantage sensitivity WEAK | | 1 | | 0 | Binomial<br>$p = .30$ | Beta<br>$\alpha = 2$<br>$\beta = 2$ or 4 |
| Vantage sensitivity STRONG | | 0 | | 0 | | |
| Differential susceptibility WEAK | | 1 | | .25/.50 | | |
| Differential susceptibility STRONG | 3 | 0 | 2 | .25/.50 | | |
| Diathesis-stress WEAK | | 1 | | 1 | | |
| Diathesis-stress STRONG | | 0 | | 1 | n = 1 or 4 | n = 1 or 4 |

Note: Small, medium, large effect sizes refers to $R^2$ = .05, .10 and .15 in the one gene and one environment case and to $R^2$ = .10, .20 and .40 in the four genes and three environments case.

## Hypothesis testing

As originally presented by Belsky et al. (2013), the competitive-confirmatory approach is primarily concerned with the nature of the gene-environment interaction (G×E), even when the G×E is not significant. Recently, Belsky and Widaman (2018) suggested using the competitive-confirmatory approach only if the F-ratio of the G×E is greater or equal to one; this strategy prevents trying to fit ill-conditioned models with near zero interaction effect, but it may not be penalizing enough to bring the rate of false positive to 5% or lower. A natural approach to minimize the presence of false positives is to additionally test for models without a G×E term. Thus, in addition to the six G×E models of interest, we also examine the following four models: (1) Intercept only, (2) gene(s) only, (3) environment(s) only, and (4) gene(s) and environment(s) only.

If any of the four models without an interaction had the lowest BIC, we classified the interaction as "no evidence of G×E". Otherwise, we classified the interaction as "differential susceptibility" if a) the weak or strong differential susceptibility models had the lowest BIC and b) the 95% interval of its estimated crossover point was within the observable bounds of the environmental score. If one of these conditions was not met, we classified the interaction based on which of the remaining four models (i.e. weak/strong vantage sensitivity and diathesis-stress models) had the lowest BIC.

For the RoS approach, we classified the pattern of interaction as reflecting "differential susceptibility" when both lower and upper bounds of the RoS were within the observable range, "vantage sensitivity" when only the upper bound of the RoS was within the observable range, "diathesis-stress" when only the lower bound of the RoS was within the observable range, or "no evidence of G×E" when neither of the bounds of the RoS were within the observable range. The lower and upper bounds were determined using the 95% confidence interval of the simple slope.

With multiple genes and environments, we found that the .05 alpha level (95% confidence intervals) lead to extremely high rates of false positives (80-97%). To remedy this issue, we decreased the alpha level until we reached a false positive rate of 15% or lower in every scenario. The resulting alpha level was .0001 (99.99% confidence intervals). Given that this is very conservative, it inevitably resulted in lower accuracy levels. However, this step was necessary in order to prevent the extremely high false positive rates.

## Results

### Single genetic and environmental variable

Results of our ability to infer the correct model from the simulations for the scenarios with a single genetic and environmental variable are shown in Figure 3. We found that the competitive-confirmatory and RoS approaches attained near-perfect accuracy in all scenarios with large sample sizes (N ≥ 1000) and large effect sizes. However, the competitive-confirmatory approach had greater accuracy than the RoS in all other scenarios.

Comparing the two scenarios in which the environmental variables are symmetric, accuracy was significantly lower when the crossover was placed at .25 compared to .50. Comparing the two scenarios in which the environmental variables were left-skewed, it can be observed that the accuracy was slightly higher when the crossover was .25 compared to .50. Note that the expected value of the environmental score is .50 and .33 when the environmental variables are Beta(2,2) and Beta(2,4) respectively. Thus, in these examples, accuracy was higher when the crossover point was near the average environmental score.

Depending on the scenario, the rates of false positives were between 0% and 4% for the competitive-confirmatory approach and between 11% and 20% for the RoS approach.

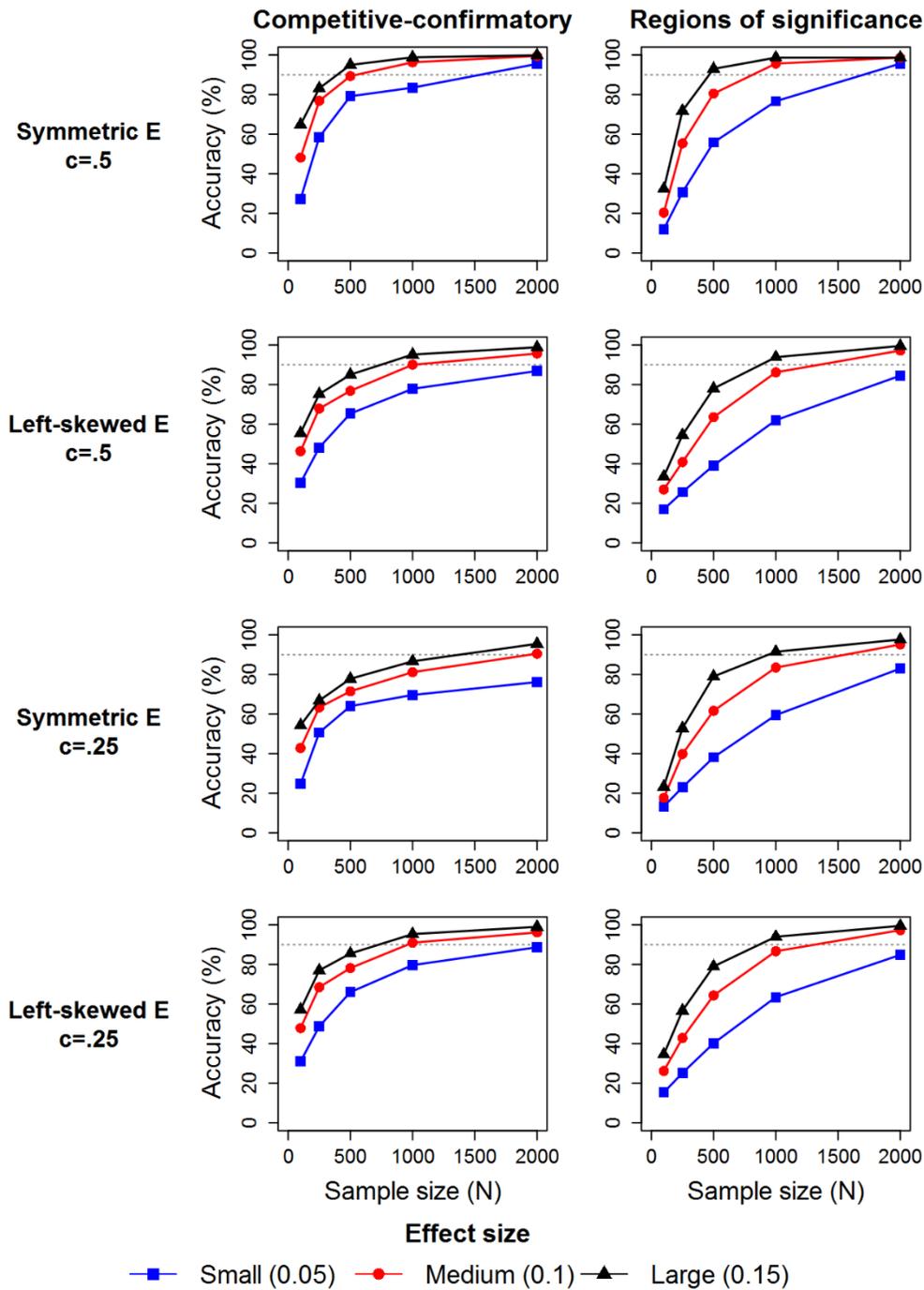

**Figure 3.** *The accuracy of the confirmatory-competitive and RoS approaches in distinguishing the type of interaction (vantage sensitivity, diathesis stress, or differential susceptibility) under various scenarios assuming a single genetic and environmental variable. For each of the six possible models (vantage sensitivity, diathesis stress, or differential susceptibility by weak/strong), 100 simulated datasets were generated. The 600 total simulated datasets were used to measure classification accuracy in the different scenarios. The scenarios varied sample size (N = 250, 500, 1000, 2000), symmetric or skewed environmental variable's distribution, crossover points at c = .25 or .50 for the differential susceptibility models and different effect*

sizes ($R^2$ = .05, .10, .15). "Symmetric E" refers to the assumption that the environmental variables are Beta(2,2) and thus symmetric, while "Left-skewed E" refers to the assumption that the environmental variables are Beta(2,4) and thus left-skewed. The variable c refers to the choice of the crossover point in the differential susceptibility models. The dotted lines represent 90% accuracy. The blue lines with square points represent $R^2$ = .05, the red lines with circle points represent $R^2$ = .10, and the black lines with triangle points represent $R^2$ = .15.

**Four genetic and three environmental variables**

Results for the scenarios with four genetic and three environmental variables are presented in Figure 4. Both the competitive-confirmatory and RoS approaches had extremely low accuracy at N = 100 and N = 250 with low/medium effect size. However, the accuracy of the competitive-confirmatory approach rapidly increased with more data points (or greater effect size); this was not the case for RoS. Importantly, the competitive-confirmatory approach attained near-perfect accuracy with large sample sizes and effect sizes, while RoS did not even for N = 2000. Similarly to the single gene, single environment scenarios, we found higher accuracy when the crossover point was proximal to the average environmental score using both approaches. Overall, both approaches had significantly lower accuracy in the multiple genes and environments scenario compared to the single gene and environment scenarios. Regarding the competitive-confirmatory models, this is explained by the fact that BIC heavily penalizes additional parameters and the difference in the number of parameters between the G×E and non-G×E models is more pronounced. In the 4 genes and 3 environments (multi-G×E) scenarios, the weak differential susceptibility model has 9 parameters compared to the one parameter of the intercept only model. In contrast, in the single gene and environment scenarios, the weak differential susceptible model has only 4 parameters, thus 3 more than the intercept only model. Regarding the RoS approach, the main reason underlying the lower accuracy rate in the multi-G×E compared to the single G×E scenarios lies in the more stringent alpha level (i.e., 0001 rather than .05, which was necessary to prevent false positives rates of ≥ 80%).

The rates of false positives were exactly 0% in all scenarios for the competitive-confirmatory approach and between 4% and 13% for the RoS approach, depending on the scenario.

When one has *a priori* knowledge about the presence of an interaction, one might want to ignore testing for non-G×E models in a competitive-confirmatory framework and use the standard .05 alpha level in RoS. We provide simulation results under those settings in Appendix D. Of note, in those settings, both approaches were markedly accurate, with the competitive-confirmatory approach having greater accuracy than the RoS approach in every scenario.

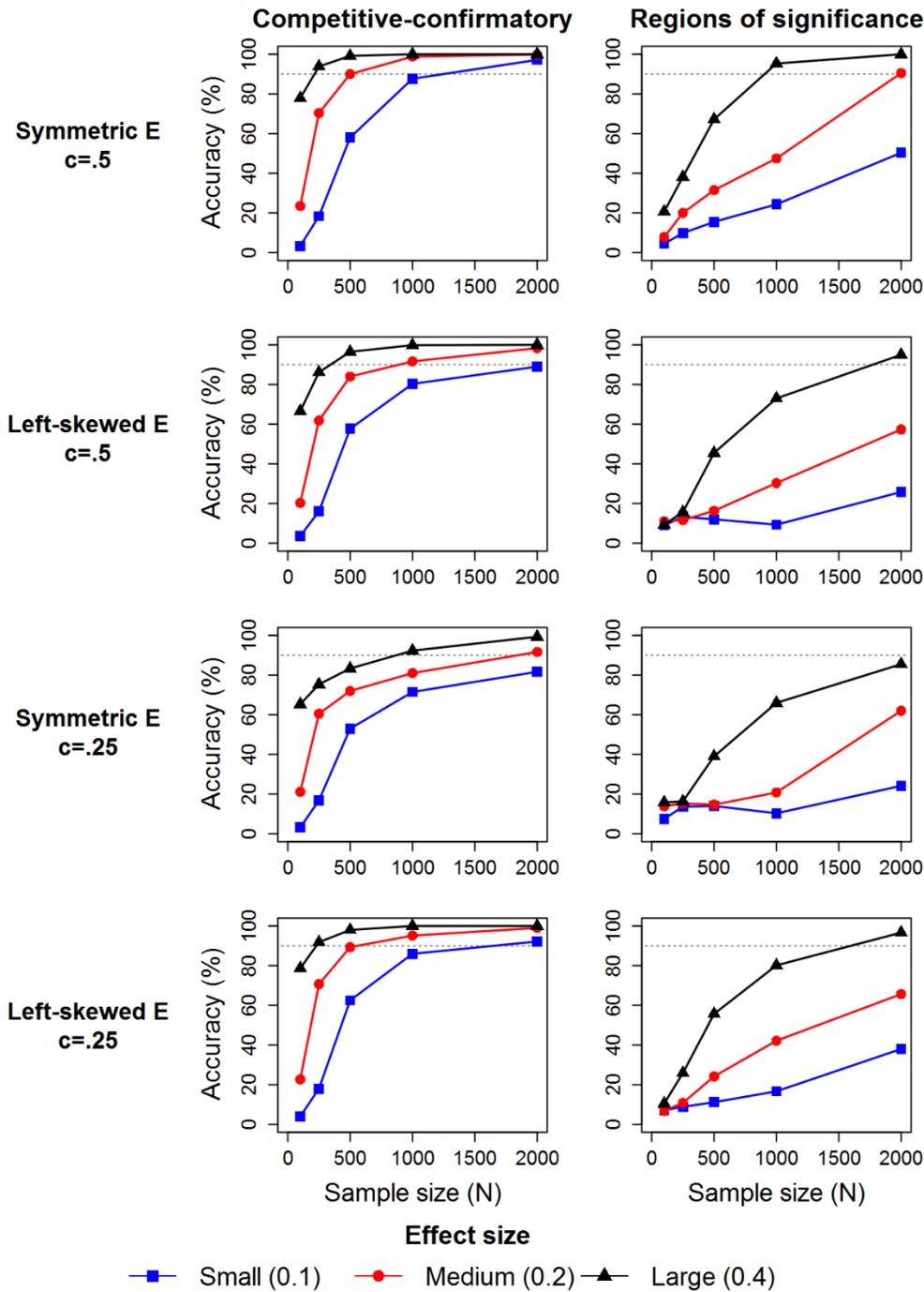

**Figure 4.** *The accuracy of the confirmatory-competitive and RoS approaches in distinguishing the type of interaction (vantage sensitivity, diathesis stress, or differential susceptibility) under various scenarios assuming four genetic and three environmental variables. For each of the six possible models (vantage sensitivity, diathesis stress, or differential susceptibility by weak/strong), 100 simulated datasets were generated. The 600 total simulated datasets were used to measure classification accuracy in the different scenarios. The scenarios varied sample size (N = 250, 500, 1000, 2000), symmetric or skewed environmental variable's distribution, crossover points at c = .25 or .50 for the differential susceptibility models and different effect*

*sizes ($R^2$ = .10, .20, .40). "Symmetric E" refers to the assumption that the environmental variables are Beta(2,2) and thus symmetric, while "Left-skewed E" refers to the assumption that the environmental variables are Beta(2,4) and thus left-skewed. The variable c refers to the choice of the crossover point in the differential susceptibility models. The dotted lines represent 90% accuracy. The blue lines with square points represent $R^2$ = .10, the red lines with circle points represent $R^2$ = .20, and the black lines with triangle points represent $R^2$ = .40.*

## Discussion

Three different conceptual models currently inform research on GXE interaction: diathesis-stress, vantage sensitivity, and differential susceptibility. Herein, we used simulations to compare two different statistical approaches; in order to do so, we also extended these concepts to situations where both the environmental and the genetic scores are derived from several variables. *Our findings indicate that the competitive-confirmatory approach performs significantly better than the RoS approach in differentiating the different models of environmental sensitivity in both single and multiple genes and environments settings*. More specifically, we observed that the competitive-confirmatory approach had good accuracy when (1) effect size was moderate and N ≥ 500, and (2) effect size was large and N ≥ 250. In practice, many G×E studies have samples of less than 500 observations and small observed effect sizes ($R^2$ < .10); consequently, our empirical results suggest that many studies are too underpowered to reliably quantify the type of interaction in a single gene and environment setting.

Our findings further indicate that the distribution of environmental factors has a noticeable impact on the accuracy of both the RoS and the competitive-confirmatory approaches. When the environment-predicting-outcome lines intersect close to the average environmental score (i.e., where the density of observations is highest), accuracy is highest; however, when the lines intersect far away from the average environmental score (i.e., where the density of observations is lowest), the accuracy tends to be lower. This implies that studies should not only report the crossover point estimate and whether it is within observable range, but also how far it is from the average environmental score.

## Computational tools

In addition to developing the LEGIT approach, we provide free and open source computational tools to perform G×E testing in R (R Development Core Team, 2017). The software we provide is available as part of the LEGIT package (cran.r-project.org/web/packages/LEGIT/).

We include a function for the RoS approach (*GxE_interaction_RoS*) and a separate function for the competitive-confirmatory approach (*GxE_interaction_test*). Although developed specifically for LEGIT models, these functions are also compatible with simple regression models by fitting a LEGIT model with only one genetic variant and one environmental measure. The software provides model fit indices, as well as the crossover point with its corresponding 95% confidence interval; it further evaluates whether that point lies within the observable range of the environmental score. Additionally, the software outputs the proportion of observations below the crossover point, a measure suggested by Roisman et al. (2012) called proportion affected (PA), and the model fit for each type of G×E (i.e., diathesis-stress, vantage sensitivity, and differential susceptibility; weak or strong). All of the six competitive-confirmatory models are available as part of the output and can be further plotted using the *plot* function.

To fit a competitive-confirmatory model, one simply needs a data frame (a standard R object used to store a dataset) comprising the outcome and possible covariates (*data*), a data frame containing the genetic variables (*genes*) and a data frame containing the environmental variables (*env*). Assuming no covariates are used, all competitive-confirmatory models can be fitted with one simple command: "*GxE_interaction_test(data=data, genes=genes, env=env, formula_no_GxE = y ~ 1)*". To add covariates, e.g., *gender* and *ses* as covariates (variables that must be part of *data*), one can change the *formula_no_GxE* option to *y ~ gender + ses*. See Figure 5 for an example output of the function. More detailed instructions on how to use the *GxE_interaction_test* function are available online (project.org/web/packages/LEGIT/vignettes/GxE_testing.html).

| | BIC | crossover | crossover 95% | | Within observable range? |
|---|---|---|---|---|---|
| Vantage sensitivity WEAK | "772.27" | "0.45" | "" | | "" |
| Differential susceptibility WEAK | "775.08" | "-0.4" | "( -0.73 / -0.07 )" | "No" | |
| Differential susceptibility STRONG | "870.68" | "2.22" | "( 1.79 / 2.64 )" | "No" | |
| Vantage sensitivity STRONG | "930.4" | "0.44" | "" | | "" |
| Diathesis-stress WEAK | "971.12" | "9.75" | "" | | "" |
| Diathesis-stress STRONG | "1123.61" | "9.74" | "" | | "" |

**Figure 5.** *Example output of the confirmatory G×E testing function from the LEGIT package.*

A few computational resources currently exist to perform RoS analyses (Hayes & Matthes, 2009; Preacher et al., 2006). However, to our knowledge, ours is the first software that directly outputs information on the pattern of G×E and with relevant details. Importantly, we also provide the code necessary to reproduce all simulations from this article and Figures 3-4 (github.com/AlexiaJM/GxETesting).

## Limitations

Although we studied a number of different scenarios, there are some unexplored scenarios that could be of interest, such as: crossover-point closer to the minimum/maximum of the environmental score (e.g. $c = .10, .75$, or $.90$), genetic variants with different proportions (rather than $p = .30$ for all), non-independent genetic variants and/or environmental factors, normally distributed or categorical environmental exposures and, as a final example, a very large number of genetic variants and environmental factors.

Given the concerns of S. Lee et al. (2015) about the non-normality of the crossover point, which is most noticeable in small samples, we also attempted bootstrapping the crossover point to obtain more robust confidence intervals. However, as the competitive-confirmatory approach does not rely directly on confidence intervals, we did not observe any significant difference in the results of our simulations. For this reason and because of the considerable computational demands of bootstrapping, we did not report the results using bootstrapping.

In randomized experiments, as done in the simulations, the environmental variables are uncorrelated with anything else, by design. However, most study design are correlational, thus the environmental variables in E could be correlated to one another or to other variables (SES, gender, etc.). To disentangle gene-by-environment effects in a correlational study design, one need to adjust for all possible confounders. Thus, one might need more observations and variables to attain the same power as a randomized experiment. This means that we could be underestimating the sample size required to attain a certain accuracy.

In the simulations with multiple genes and environments, we assumed that one knows which genes and environments to include a priori. However, in practice, one does not know exactly which genes and environments to use. Thus, one often use some form of variable selection technique to determine which subset of genes and environments to use. This may results in a certain amount of multiple testing, depending on the variable selection method used. We did not study the impact of variable selection on the accuracy of the competitive-confirmatory and RoS approaches.

Questions have been raised about the evolutionary plausibility of vantage sensitivity (i.e., differential responsiveness to just positive experiences and exposures)(Bakermans-Kranenburg & van IJzendoorn, 2015). We agree that if there were no costs associated with vantage sensitivity, in contrast to the notion of differential susceptibility (i.e., being affected by both positive and negative experiences/exposures: "for better and for worse"), then the expectation would be that genetic variants making some individuals highly responsive to just enriching experiences would, over time, go to fixation, spreading to all individuals because of its beneficial consequences. Once fixated, there would be no variation in susceptibility to only supportive and enriching experiences as a function of these variants. However, if such fixation is not complete, variation in susceptibility to only positive experiences is plausible.

## Conclusion

In this paper, we showed how to adapt the theory and computation of the competitive-confirmatory and RoS approaches to the study of G×E interaction to multi-genes/multi-environments settings using LEGIT models. Furthermore, through careful simulation analyses, we demonstrated that the competitive-confirmatory approach performs significantly better than the RoS approach. We then showed that accuracy in the multi-gene/multi-environment setting was lower than in the single-gene/singe-environment setting, however, the competitive-confirmatory approach maintained good accuracy at large sample sizes and large effect sizes (N ≥ 250 with large effect size, N ≥ 500 with medium effect, and N ≥ 1000 with small effect size), which was not the case for RoS. Given these results, *we strongly advise researchers to switch from the RoS approach to the competitive-confirmatory approach when testing the form of the interaction*, considering its overall greater accuracy in distinguishing diathesis-stress, vantage sensitivity, and differentiability susceptibility. We believe that considering multiple genes and environments in a single model is of great importance given the vast amount of non-replicable results, generally arising from models with very small effect sizes. LEGIT is a user-friendly and freely accessible R package which will aid researchers in implementing these recommendations. Currently, the LEGIT approach is not readily applicable to genome-wide association study (GWAS) data given overparametrization issues (p >> n) and the high risk of multi-collinearity resulting from the very large correlation between many genetic variants; in the future, our approach could be adapted for genome-wide data.

## Appendix A: Adapting the alternating optimization algorithm

To adapt the alternating optimization algorithm to competitive-confirmatory models, the negative intercept (the crossover point $c$) must be excluded from the constraint that all environmental weights sum to 1 in absolute values (i.e., $-c + \sum_{l=1}^{s} |q_l| = 1$ cannot be enforced); this is because the crossover point must be free. While the original approach by Jolicoeur-Martineau et al. (2017) was guaranteed to converge to a local minimum, not enforcing a constraint on $c$ means that the model can diverge if the starting point of the crossover is not close enough to its true value or if the true value is very close to -∞ or ∞. This can be prevented by fitting a G×E model with the regular formulation (equation 6), calculating the value of the cross-over point $c = -\dfrac{\hat{\beta}_g}{\hat{\beta}_{eg}}$ and refitting the model with the formulation which includes the cross-over point as variable (equation 9) using $c$ as starting point. If the crossover point still diverges toward -∞ or

∞, it generally indicates that the interaction effect is too small to estimate and is thus essentially nil (Widaman et al., 2012). Therefore, Belsky and Widaman (2018) suggest testing an interaction only when the magnitude of the F-ratio of the interaction is larger than 1. Fortunately, in models that include latent genetic and environmental scores, this issue occurs infrequently, because the inclusion of multiple genetic and environmental variables in the model tends to considerably increase the effect size of the interaction (Jolicoeur-Martineau et al., 2017).

### Appendix B: Special considerations and recommendations

In addition to the small change in the alternating optimization algorithm, adapting the theory and testing of G×E to multiple genes and environments requires a few special considerations which need to be taken into account, as they influence how one should construct the competitive-confirmatory models and how the simulations will be set up.

Firstly, the transition from a binary genetic variant to an approximately continuous latent genetic score affects the interpretation of weak and strong competitive-confirmatory models. In cases of a single gene and environment, the interpretation is that individuals who possess no sensitivity gene variants (with a genetic score of 0) are completely non-susceptible to their environment in the strong models ($\beta_e = 0$) and somewhat susceptible in the weak model ($\beta_e \neq 0$). However, as the number of included genetic variants increases, the probability of an individual possessing no sensitivity gene variants (genetic sensitivity score of 0) approaches zero.

Assuming that we have k independent binary genetic variants with different frequencies, i.e. $g_i \sim \text{Bernoulli}(\theta_i)$ for $i = 1, \ldots, k$

We have

$$\lim_{k \to \infty} P(g_1 + g_2 + \ldots + g_k = 0) = \lim_{k \to \infty} P(g_1 = 0) P(g_2 = 0) \ldots P(g_k = 0)$$
$$= \lim_{k \to \infty} (1 - \theta_1)(1 - \theta_2) \ldots (1 - \theta_k)$$
$$= 0$$

The above means that, when including multiple genetic variants, extremely few (if any) individuals are expected to be completely non-susceptible to their environment. Consequently, weak models cannot be differentiated from strong models based on whether there are individuals that are completely non-susceptible. The only difference between weak and strong models, then, is that the slope of the environment-predicting-outcome regression line starts (with a genetic score of 0) at 0 for strong models and at $\beta_e$ for weak models. In practice, distinguishing between weak and strong models is difficult, especially when sample size is small. Thus, although we fit weak and strong models using the competitive-confirmatory approach, we do not consider the difference between weak and strong meaningful for interpretation nor classification purposes. Accordingly, our analyses of simulated data attempt only to classify the general pattern of interaction (i.e. whether it represents diathesis-stress, differential susceptibility, or vantage sensitivity) but not whether the pattern of interaction is weak or strong. Given that the RoS approach cannot naturally distinguish weak and strong models, not trying to determine whether an interaction is weak or strong using the competitive-confirmatory approach means that we can directly compare both approaches.

Secondly, the observable range of the environmental score is unknown a priori (before fitting the model) as the weights of the environmental variables have to be estimated first and the variables included in the environmental score often have different ranges. This not only makes interpretation of the environmental score difficult, but it also makes it difficult to properly select the fixed value of the cross-over point to be used in the vantage sensitivity and diathesis-stress models. To help with interpretation, we recommend rescaling all environmental variables beforehand using a method, such as POMP coding (Cohen, Cohen, Aiken, & West, 1999); this approach rescales all variables to lie in the range [0,100] where 0 is the minimum and 100 the maximum value. Certainly, this step can be ignored when all environmental variables already have the same range of values (e.g., when using the same questionnaire at different time points).

However, even if all environmental are scaled to the same range of observable values, a problem remains; the observed score range will be smaller than the observed range of the environmental variables (e.g. [0,100] when using POMP coding) unless individuals exist in the sample who carry only minimum values for all environmental measures, and conversely individuals with only maximum values for all environmental measures. This issue is encountered very often in small samples and when the environmental variables have unbounded distributions (e.g., normal distribution) or bounded but heavily skewed distributions. Nevertheless, in order to be able to construct the diathesis-stress and vantage sensitivity models, we need to fix the crossover point at, or near, the maximum and minimum observable environmental quality score, respectively.

To account for this issue, we recommend fixing the crossover point to the expected maximum or minimum of the environmental quality score, which is the environmental score that would be obtained when one has the highest or lowest possible values on all environmental variables. With POMP coding, this corresponds to simply setting the crossover point to 100 for diathesis-stress and to 0 for vantage sensitivity. We found that this approach markedly improved simulation results in small sample settings as opposed to using the observable minimum and maximum. We justify this approach with an example and through geometric intuition.

As an example, assuming that we have a few environmental variables in the range [0, 100] and that the true pattern of interaction is differential susceptibility with $c = 25$. Let say that, by chance, the sample observed environmental scores have a restricted range such as [10, 90] instead of the actual maximum possible range [0, 100]. If we set the crossover point of the vantage sensitivity models at the observed minimum ($c_{low} = 10$) rather than the theoretically possible minimum ($c_{low} = 0$), the probability of misclassification will increase. This can be explained by the fact that the closer we fix the cross-over point of the vantage sensitivity models to the real crossover of the differential susceptibility model, the more difficult it will be to differentiate the two models. Similarly, if the true pattern of interaction is diathesis-stress ($c_{low} = 100$) and we set the crossover point of the diathesis-stress models at the observed minimum ($c_{low} = 90$) rather than the theoretically possible maximum, the probability of misclassification will increase.

Figure 6 shows a geometric intuition to why it is preferable to fix the cross-over point of the vantage sensitivity and diathesis-stress models at values slightly outside the range of possible values than at values slightly inside the range of possible values. Thereby, using the expected maximum or minimum environmental score (i.e., the environmental score that would be obtained when one has the highest or lowest possible values on all environmental variables) is preferable to using the observed maximum or minimum. A more thorough mathematical explanation of this intuitive argument is available in Appendix C.

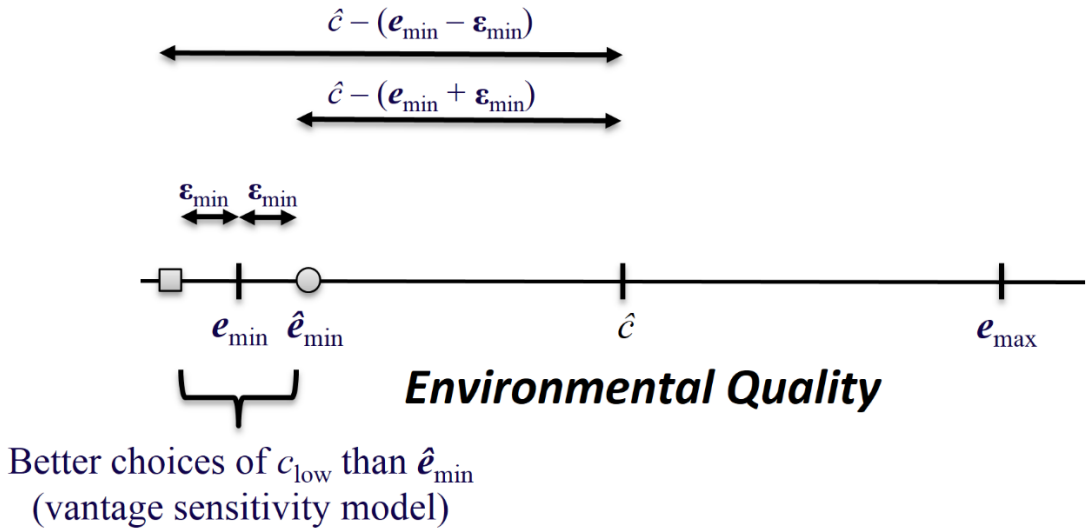

**Figure 6.** Assuming $\hat{c}$ is the estimed crossover point in the differential susceptibility model, $e_{min}$ and $e_{max}$ are the minimum/maximum of the environmental quality, $\hat{e}_{min}$ is the observable minimum of the environmental quality and $\varepsilon_{min} = \hat{e}_{min} - e_{min}$. We show two extreme choices of the fixed crossover point in the vantage sensitivity model ($c_{low}$) with the same distance to $e_{min}$, one at $\hat{e}_{min}$ (the square) and one at $\hat{e}_{min} - 2\varepsilon_{min}$ (the circle). To better classify the pattern of interaction, we want the vantage sensitivity model to be differentiable from the differential susceptibility model, thus, intuitively, we want to maximize the distance between $c_{low}$ and $\hat{c}$. Clearly, any choice of $c_{low}$ between the square point ($\hat{e}_{min} - 2\varepsilon_{min}$) and the circle point ($\hat{e}_{min}$) is better than $\hat{e}_{min}$. We conclude that choosing $c_{low}$ that is slightly too small will likely lead to lower risk of misclassification than choosing $c_{low}$ that is slightly too large.

### Appendix C: Choice of crossover point in vantage sensitivity and diathesis-stress models

To minimize the probability of misclassification, we want that 1) the true (unobserved) minimum ($e_{min}$) and maximum ($e_{max}$) of the environmental score are close enough to the fixed crossover point of the vantage sensitivity models ($c_{low}$) and of the diathesis-stress models ($c_{high}$) so that these models correctly represent vantage sensitivity and diathesis-stress respectively and 2) the vantage sensitivity and diathesis-stress models are as far as possible from the differential susceptibility models in terms of fit. Assuming that $\hat{c}$ is the estimated crossover point from the differential susceptibility models, we can formulate this in the following way:

- Choose the fixed crossover point of the vantage sensitivity models ($c_{low}$) such that
  1) the distance between $c_{low}$ and $e_{min}$ is minimized
  2) the distance between $c_{low}$ and $\hat{c}$ is maximized

- Choose the fixed crossover point of the diathesis-stress models ($c_{high}$) such that
    1) the distance between $c_{high}$ and $e_{max}$ is minimized
    2) the distance between $c_{high}$ and $\hat{c}$ is maximized

This is represented visually in Figure 5. Choosing $c_{low} = e_{min}$ and $c_{high} = e_{max}$ would be ideal but we do not know $e_{min}$ and $e_{max}$. However, we know the observable minimum and maximum of the environmental score in the sample ($\hat{e}_{min}$ and $\hat{e}_{max}$) and we know that the $e_{min} \leq \hat{e}_{min}$ and $e_{max} \geq \hat{e}_{max}$. Assuming the distance between the observed and true minimum of the environmental score ($\hat{e}_{min} - e_{min}$) is $\varepsilon_{min}$ and the distance between the observed and true maximum of the environmental score ($\hat{e}_{max} - e_{max}$) is $\varepsilon_{max}$. Then, any choice of $c_{low}$ such that $c_{low} < \hat{e}_{min}$ and $c_{low} > \hat{e}_{min} - 2\varepsilon_{min}$ will be better, with respect to the two objectives above, than choosing $c_{low} = \hat{e}_{min}$. Similarly, any choice of $c_{high}$ such that $c_{high} < \hat{e}_{max}$ and $c_{high} > \hat{e}_{max} + 2\varepsilon_{max}$ will be better, with respect to the two objectives above, than choosing $c_{high} = \hat{e}_{max}$. Overall, this means that to reduce the risk of misclassification when do not know the true minimum and maximum of the environmental score, it is preferable to fix the cross-over point of the vantage sensitivity and diathesis-stress models at values slightly outside the range of possible values than at values slightly inside the range of possible values.

Note that by choosing the theoretical minimum and maximum as the fixed value of the crossover point in the vantage sensitivity and diathesis-stress models respectively, we are not guaranteed that $c_{low} > \hat{e}_{min} - 2\varepsilon_{min}$ and $c_{high} > \hat{e}_{max} + 2\varepsilon_{max}$. However, unless one's data is extremely skewed or that certain combinations of values are impossible (e.g. if there are two environmental variables and we have that one is happiness and the other is sadness, then it is impossible to have high or low values in both variables), these assumptions are likely to apply. We encourage researchers to act with sound judgment and decide if these assumptions are reasonable or not in their own analysis; otherwise, simply using the observable minimum or maximum as the crossover point for the vantage sensitivity and diathesis-stress could be preferable.

## Appendix C: Multiple genes and environments scenarios when not testing for non-G×E models in competitive-confirmatory and using alpha level of .05 in RoS

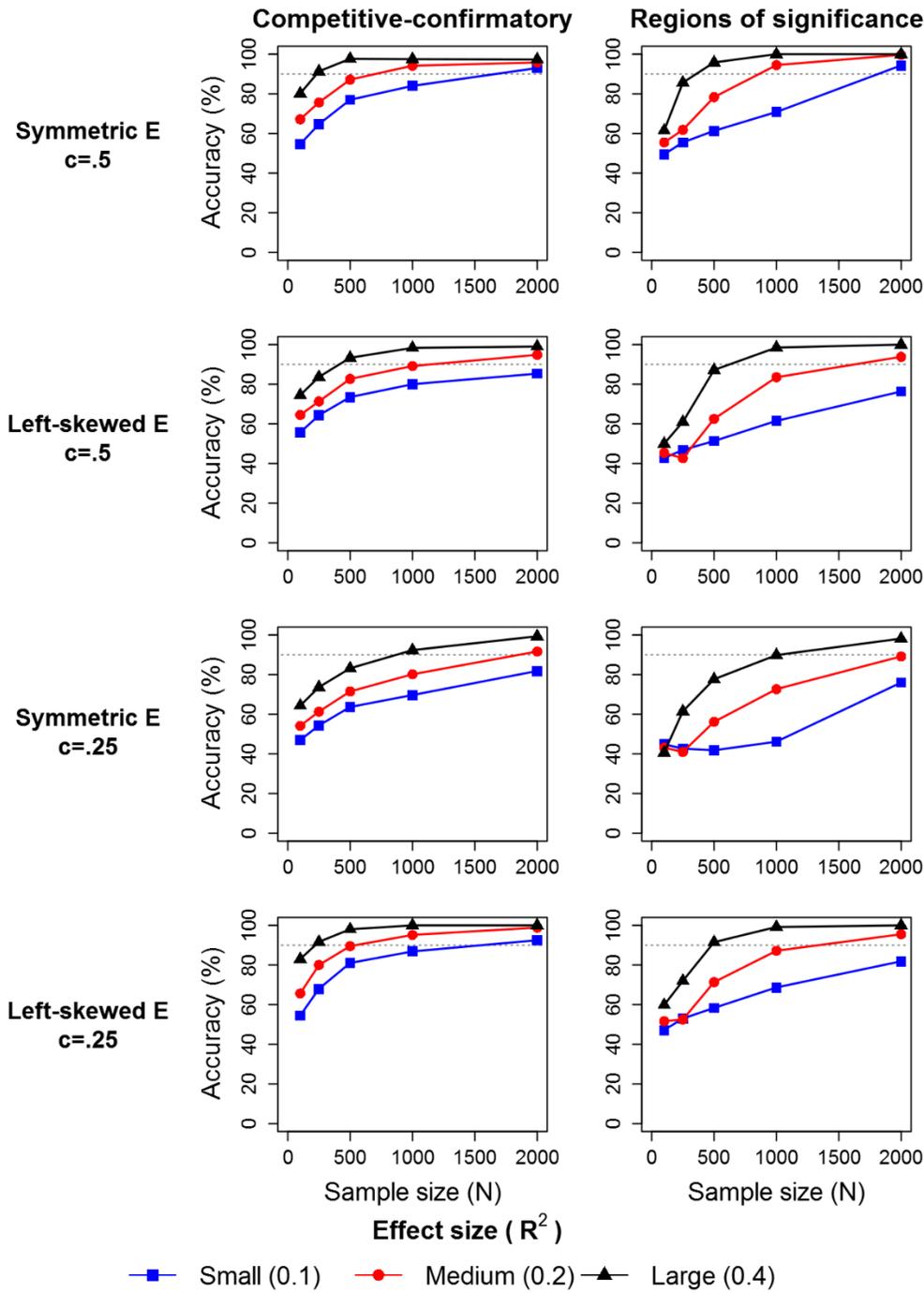

**Figure 7.** Reproducing Figure 4 when not testing for non-G×E models in competitive-confirmatory and using alpha level of .05 in RoS (i.e. when assuming that we have an interaction and not taking any step to prevent false positive).